\documentclass[aps,preprint]{revtex4}
\usepackage{graphicx}
\usepackage[dvips]{color}
\begin{document}

\title{Heat cost of parametric generation of microwave squeezed states}

\author{A.M. Zagoskin$^{1,2}$}
\author{E. Il'ichev$^3$}
\author{Franco Nori$^{2,4}$}
\affiliation{%
(1) Department of Physics, Loughborough University, Loughborough, Leics LE11 3TU, UK}
\affiliation{(2) Advanced Science Institute, RIKEN, Wako-shi, Saitama 351-0198, Japan}
\affiliation{(3)
Institute of Photonic Technology, P.O. Box 100239, D-07702 Jena,
Germany}
\affiliation{(4) Physics Department, The University of Michigan, Ann Arbor, MI 48109-1040, USA}

\date{\today}

\begin{abstract}
In parametric systems, squeezed states of radiation can be generated via extra work done by external sources. This eventually increases the entropy of the system despite the fact that squeezing is reversible. We investigate the entropy increase due to squeezing and show that it is quadratic in the squeezing rate and may become important in the repeated operation of tunable oscillators (quantum buses) used to connect qubits in various proposed schemes for quantum computing.  
\end{abstract}

\maketitle

Parametric devices using the nonlinear inductance and natural protection from decoherence of Josephson junctions are considered as prospective elements of quantum circuits \cite{Nation2011}. In particular, these have been proposed and implemented as quantum buses, providing tunable coupling between solid state qubits (see, e.g., \cite{Makhlin2001,Blais2003,Wendin2006,You2011,Zagoskin2011}). On the other hand, parametric Josephson devices can produce squeezed states of microwave radiation \cite{Everitt2004,Hu1995,Moon2005,Ojanen2007,Zagoskin2008}. Such experiments were performed in the 1980s \cite{Yurke1988,Yurke1989} and more recently \cite{Castellanos-Beltran2008}. 

In this paper we investigate the generation of squeezed states during the operation of a quantum bus and estimate their additional contribution to the entropy of the system and associated heat production. We will see that this contribution  is quadratic in the squeezing rate  and should be taken into account if the system operates near its limit of efficiency determined by the Landauer erasure principle (see, e.g., \cite{Plenio2001,Maruyama2009}).

A quantum bus can be considered as a harmonic oscillator with tunable frequency $\omega(t)$; tuning it in and out of resonance with qubits allows to manipulate their states   (e.g., by performing two-qubit gates). The changes of $\omega(t)$ should be fast compared to the characteristic frequencies of qubits and the bus, i.e., $\dot{\omega}/\omega \gg \omega$; the decoherence rate in the circuit, however, must be much smaller than $\omega$. But these are the very conditions which lead to the generation of squeezed states. Indeed, a coherent state of a harmonic oscillator   is squeezed by a sudden change of the oscillator frequency~\cite{Graham1987}. Recall that   a coherent state is a state with equal and minimal
uncertainties:  $ \langle \Delta P^2 \rangle = \langle\Delta Q^2
\rangle = 1/2$, where $P = (a - a^{\dag})/i\sqrt{2}$ and $Q =  (a + a^{\dag})/\sqrt{2}$ are dimensionless momentum and position expressed through bosonic creation and annihilation operators.

 It is convenient to describe oscillator states by their Wigner functions (see, e.g., \cite{Gardiner2004}):  
\begin{equation}
W(\alpha,\alpha^*) = \frac{1}{2\pi^2} \int\: d\lambda\:d\lambda^* \: e^{(-\lambda \alpha^* + \lambda^* \alpha)} \;{\tt tr} \left[ e^{(\lambda a^{\dag} - \lambda^* a)} \rho\right],
\label{eq:Wigner-def}
\end{equation}
which have the advantage of reducing to classical distribution functions in the classical limit. Here $\rho$ is the density matrix of the system. For a coherent state, the Wigner function is Gaussian, $$W_{\alpha_0}(\alpha, \alpha^*) = \frac{2}{\pi} \exp[-2|\alpha - \alpha_0|^2].$$
 The  amplitude and phase of the complex parameter $\alpha_0$ (which in the Schr\"{o}dinger representation behaves as $\alpha_0(t) = \alpha_0(0) \exp[-i\omega t]$) describe the classical limit of the oscillator state. A squeezed state will have instead the Wigner function 
\begin{equation}
W(\alpha, \alpha^*) = \frac{2}{\pi} \exp\left\{-2 s \left[ (x-x_0)\cos\theta + (y-y_0)\sin\theta  \right]^2 - \frac{2}{s}\left[ (y - y_0)\cos\theta - (x-x_0)\sin\theta\right]^2\right\}.
\label{eq:squeezed-Wigner-def}
\end{equation}
Here $x = {\tt Re}\; \alpha, y = {\tt Im}\; \alpha$, $s$ is the squeezing parameter, and $\theta$ determines the direction of the squeezing axis. Similarly, the thermal state $$ W_{\rm th}(\alpha, \alpha^*) = \frac{2/\pi}{1+2\bar{n}} \exp[-2|\alpha|^2/(1+2\bar{n})],$$ characterized by the average photon number $\bar{n} = \left\{\exp\left[\omega/T\right]-1\right\}^{-1}$, can be squeezed to \cite{Kim1989} 
\begin{equation}
W(\alpha, \alpha^*) = \frac{2/\pi}{1+2\bar{n}} \exp\left\{-\frac{2}{1+2\bar{n}} \left[ s \left( x\cos\theta + y\sin\theta  \right)^2 + \frac{1}{s}\left( y\cos\theta - x\sin\theta\right)^2\right]\right\}.
\label{eq:squeezed-Wigner-thermal}
\end{equation}

After being introduced in Ref.~\cite{Graham1987}, squeezing the oscillator states by a sudden change of oscillator frequency was considered in a number of papers \cite{Agarwal1991,Janszky1992,Abdalla1993,Kiss1994,Averbukh1994,Hu1995,Zagoskin2008}. In particular, it was shown \cite{Janszky1992,Averbukh1994,Zagoskin2008} that, in the absence of decoherence, repeated abrupt small changes of the oscillator frequency can produce arbitrarily large squeezing. Here we will mainly follow the approach of \cite{Zagoskin2008}.

Consider an arbitrary time dependence of the system frequency $\omega(t)$. Let us also denote the creation/annihilation operators belonging to the Fock state of an oscillator with  the frequency $\omega(t=0)$ by $a_0, \:a^{\dag}_0$, and those corresponding to $\omega(t)$, by $a_{\omega}, \:a^{\dag}_{\omega}$. The Hamiltonian keeps its standard form,
\begin{equation}
H(t) = \hbar\omega(t) \left(a^{\dag}_{\omega}a_{\omega} + \frac{1}{2}\right),
\label{eq:4-time-dependent-H}
\end{equation}
and  the commutation relations between $a_{\omega}, \:a^{\dag}_{\omega}$ hold, if the old and new operators are related via a Bogoliubov transformation 
\begin{equation}
a_{\omega} = \frac{\left[\omega(t)+\omega(0)\right]a_0 - \left[\omega(t)-\omega(0)\right]a^{\dag}_0}{2\sqrt{\omega(t)\omega(0)}},
\label{eq:4-Bogoliubov-squeezed-2}
\end{equation}
 which can be explicitly written as  \cite{Graham1987}
\begin{equation}
a_0 = V^{\dag}(t) a_{\omega} V(t); \:\: a^{\dag}_0 = V^{\dag}(t) a^{\dag}_{\omega} V(t)
\label{eq:4-unitary-squeezed-V}
\end{equation}
where 
\begin{equation}
V(t) = \exp\left\{-\frac{1}{4}\left[ \ln \frac{\omega(0)}{\omega(t)} \right] \left[a_{\omega}^2 - (a^{\dag}_{\omega})^2 \right]\right\}; \:\: V^{\dag}(t) = \exp\left\{\frac{1}{4}\left[ \ln \frac{\omega(0)}{\omega(t)} \right] \left[a_{\omega}^2 - (a^{\dag}_{\omega})^2 \right]\right\}.
 \label{eq:4-V-transform}
\end{equation}
In order to work in the initial Fock space at $t=0$, we apply  the transformation $V^{-1}(t) = V^{\dag}(t)$, which  gives
\begin{eqnarray}
H(t) \to \tilde{H}(t) = V(t)H(t)V^{\dag}(t) - i\hbar V(t) \frac{\partial}{\partial t}V^{\dag}(t) = \nonumber\\
\hbar\omega(t) \left(a_0^{\dag}a_0 + \frac{1}{2} \right) + i \hbar \frac{\dot{\omega}(t)}{\omega(t)} \left[a_{0}^2 - (a^{\dag}_{0})^2 \right].
\label{eq:4-unsqueezed-H}
\end{eqnarray}
Hereafter we correct an error made in \cite{Zagoskin2008}, where the transformation (\ref{eq:4-unsqueezed-H}) was effectively applied twice in the same direction, which quantitatively (but not qualitatively) affected the results. For the Wigner function, with $\alpha,\: \alpha^*$ always referring to the coherent states in the same Fock space (at $t=0$), we find  the master equation (\cite{Zagoskin2008}, mutatis mutandis)
\begin{equation}
\frac{\partial}{\partial t}W(\alpha, \alpha^*, t) = 2\omega(t) {\tt Im} \left(\alpha^* \frac{\partial}{\partial \alpha^*}\right) W(\alpha, \alpha^*, t) + \frac{\partial \ln \omega(t)}{\partial t} {\tt Re} \left(\alpha \frac{\partial}{\partial \alpha^*}\right) W(\alpha, \alpha^*, t),
\label{eq:4-squeezing-W-complex}
\end{equation}
or 
\begin{equation}
\frac{\partial}{\partial t}W(x, y, t) = \omega(t)   \left(x\frac{\partial}{\partial y} - y\frac{\partial}{\partial x}\right) W(x, y, t) + \frac{1}{2}\frac{\partial \ln \omega(t)}{\partial t}   \left(x \frac{\partial}{\partial x} - y \frac{\partial}{\partial y}\right) W(x, y, t).
\label{eq:4-squeezing-W-real}
\end{equation}
We did not include the diffusive terms, which describe decoherence (including relaxation), thus assuming that the system maintains coherence over many cycles of frequency-switching and of its free evolution with frequency $\omega(t)$.  
Equation~(\ref{eq:4-squeezing-W-real}) is a  first-order linear equation and can be solved by the method of characteristics. The  characteristic equations are
\begin{eqnarray}
\frac{dx}{dt} = \frac{\dot{\omega}}{2\omega} x - \omega y; \:\:
\frac{dy}{dt} =  \omega x - \frac{\dot{\omega}}{2\omega} y.
\label{eq:4-characteristics}
\end{eqnarray}
In the limit of fast frequency change, when the $\dot{\omega}$-terms dominate, these equations lead to squeezing of the quantum state:
\begin{equation}
\frac{d \ln x}{dt} \approx  \frac{d \ln \sqrt{\omega}}{dt};\:\:\: \frac{d \ln y}{dt} \approx  -\frac{d \ln \sqrt{\omega}}{dt}, 
\label{eq:characteristic-squeezing}
\end{equation}
so that $W(x, y, \Delta t) \approx W(\sqrt{s} x, \frac{1}{\sqrt{s}} y, 0)$, with  squeezing parameter $s = \omega(\Delta t)/\omega(0)$, and $\Delta t$ the duration of the fast frequency change [cf. Eqs. (\ref{eq:squeezed-Wigner-def}, \ref{eq:squeezed-Wigner-thermal})]. In the quasistatic limit, Eq.~(\ref{eq:4-characteristics}) simply describes the rotation of the Wigner function as a whole: slow changes of parameters do not produce any squeezing, as expected.  

As we mentioned above, in the absence of decoherence, periodic changes of the oscillator frequency can produce an arbitrarily high degree of squeezing, even if the difference betweeen the two limiting values, $\omega_0$ and $\omega_1$, is arbitrarily small. It is only necessary that at least one of the transitions ($\omega_0 \to \omega_1$ or $\omega_0 \to \omega_1$) is fast on the scale of $\omega$, and these transitions are tuned to the phase of the oscillator. This may also take place during the operation of a quantum bus, which would normally switch fast on- and off-resonance with the qubits coupled to it. As we shall see, there are  entropy costs associated with the production of squeezed states. Therefore this side-effect should be avoided in the operation of quantum buses.

Let us return to the characteristics equation (\ref{eq:4-characteristics}). Without loss of generality, we can consider two regimes of oscillator frequency switches $\omega_0 \leftrightarrow \omega_1$: the ``ratchet" regime (when one transition, e.g., $\omega_0 \to \omega_1$, is fast, and the opposite one is slow \cite{Averbukh1994,Zagoskin2008}), and the ``seesaw" regime (when both transitions are fast \cite{Janszky1992}). In either case, if two consecutive fast switches occur at the moments $t_{n-1}$ and $t_n$, the point on a characteristic will evolve according to 
\begin{eqnarray}
\left( \begin{array}{l} x \\ y \end{array}\right)_{t_n+0} = \left( \begin{array}{ll} \sqrt{s_n} \cos \theta_{n,n-1} &  \sqrt{s_n} \sin \theta_{n,n-1} \\
-\frac{1}{\sqrt{s_n}} \sin \theta_{n,n-1} &  \frac{1}{\sqrt{s_n}} \cos \theta_{n,n-1} \end{array} \right) 
\left( \begin{array}{l} x \\ y \end{array}\right)_{t_{n-1}+0} \equiv \Lambda_n \left( \begin{array}{l} x \\ y \end{array}\right)_{t_{n-1}+0}.
\label{eq:transformatrix}
\end{eqnarray}
Here $\theta_{n,n-1}$ is the phase angle accumulated during the period of slow evolution, and $s_n$ is the squeezing achieved at the $n$th step. In the ``ratchet" regime: $s_n = s = \omega_1/\omega_0$ (or vice versa), while in the ``seesaw" case: $s_{2n} = 1/s_{2n+1} = s$. Obviously, $\det\Lambda_n = 1$.   After two consecutive switches
\begin{eqnarray}
\left(\Lambda_n\Lambda_{n-1}\right)^{\rm r} =  \nonumber \\  \!\!\!\!\!\!\!\!\!\!\!\!\!\!\!\! \left( \begin{array}{ll} (s-1) \cos \theta_{n,n-1} \cos \theta_{n-1,n-2} + \cos \theta_{n,n-2}&  (s-1) \cos \theta_{n,n-1} \sin \theta_{n-1,n-2} + \sin \theta_{n,n-2} \\   -\left( \frac{1}{s}-1 \right) \cos \theta_{n,n-1} \sin \theta_{n-1,n-2} - \sin \theta_{n,n-2}&  \left( \frac{1}{s} -1\right) \cos \theta_{n,n-1} \cos \theta_{n-1,n-2} + \cos \theta_{n,n-2}  \end{array} \right)
\label{eq:transformatrix=ratchet}
\end{eqnarray}
and
\begin{eqnarray}
\left(\Lambda_n\Lambda_{n-1}\right)^{\rm s} = \nonumber \\
 \!\!\!\!\!\!\!\!\!\!\!\!\!\!\!\!  \left( \begin{array}{ll} -(s-1) \sin \theta_{n,n-1} \sin \theta_{n-1,n-2} + \cos \theta_{n,n-2}&  \:\:\:\:\:(s-1) \sin \theta_{n,n-1} \cos \theta_{n-1,n-2} + \sin \theta_{n,n-2} \\   -\left( \frac{1}{s}-1 \right) \sin \theta_{n,n-1} \cos \theta_{n-1,n-2} - \sin \theta_{n,n-2}&  -\left( \frac{1}{s} -1\right) \sin \theta_{n,n-1} \sin \theta_{n-1,n-2} + \cos \theta_{n,n-2}  \end{array} \right).
\label{eq:transformatrix=seesaw}
\end{eqnarray}
Here $\theta_{n,n-2} = \theta_{n,n-1} + \theta_{n-1,n-2}$. In the case of periodic switchings, $\theta_{n+1,n} = \Theta$, this reduces to
\begin{eqnarray}
\left(\Lambda_n\Lambda_{n-1}\right)^{\rm r} =  \left( \begin{array}{ll} (s-1) \cos^2 \Theta + \cos 2\Theta &   \:\:\:\:\:\:\:\:\:\:\:\:\:\:\:\:\:\:\:\:\:\:\:\:\:\:\:\frac{s+1}{2} \sin 2\Theta  \\   -\frac{ 1/s + 1}{2} \sin 2\Theta &  \:\:\:\:\left(\frac{1}{s} -1\right) \cos^2 \Theta + \cos 2\Theta \end{array} \right)
\label{eq:transformatrix=ratchet-regular}
\end{eqnarray}
and
\begin{eqnarray}
\left(\Lambda_n\Lambda_{n-1}\right)^{\rm s} =  \left( \begin{array}{ll} -(s-1) \sin^2 \Theta + \cos 2\Theta &  \:\:\:\:\:\:\:\:\:\:\:\:\:\:\:\:\:\:\:\:\:\:\:\:\:\:\:\:\:\:\:\:\:\frac{s+1}{2} \sin 2\Theta  \\   -\frac{ 1/s + 1}{2} \sin 2\Theta &\:\:\:  -\left(\frac{1}{s} -1\right) \sin^2 \Theta + \cos 2\Theta \end{array} \right).
\label{eq:transformatrix=seesaw-regular}
\end{eqnarray}
The resonance conditions, $\Theta^{\rm r} = \pi q$ and $\Theta^{\rm s} = \pi (q+1/2)$, when either (\ref{eq:transformatrix=ratchet-regular}) or (\ref{eq:transformatrix=seesaw-regular}) diagonalize, ensure that after $2N$ switchings the state will be squeezed exponentially to an arbitrary degree,
\begin{eqnarray}
\left( \begin{array}{l} x \\ y \end{array}\right) = \left( \begin{array}{ll} \pm s &  0  \\    0 &  \pm \frac{1}{s} \end{array} \right)^N \left( \begin{array}{l} x_0 \\ y_0 \end{array}\right) = \pm \left( \begin{array}{l} s^N x_0 \\ s^{-N} y_0 \end{array}\right),
\label{eq:ultimate squeezing}
\end{eqnarray}
no matter how small $(s-1)$. For finite values of $(s-1)$, an exact resonance is not necessary. For example, in the ratchet case  a runaway squeezing will happen if \cite{Averbukh1994}
\begin{equation}
s > s_c = \frac{1 + |\sin\Theta|}{|\cos\Theta|}.
\label{eq:critical-s}
\end{equation}
Deviations from periodic switching in the operation of a quantum bus, fluctuations of circuit parameters, and eventually relaxation and decoherence in the system will limit the actual degree of squeezing. Nevertheless, the very operation of a quantum bus will tend to produce squeezed states.

Squeezing is in itself a reversible process. Nevertheless, in the presence of decoherence it will lead to an increase of the system's entropy, which can be associated with  internal friction \cite{Feldmann2006}. To be specific, consider the so-called ``energy entropy" \cite{Feldmann2004,Feldmann2006}
\begin{equation}
S_E[\rho] \equiv -\sum_n p_n \ln p_n \geq S[\rho] \equiv -{\tt tr} \left[\rho \ln \rho\right],
\label{eq:energy-entropy}
\end{equation} 
where $\rho$ is the density matrix, $\{p_n\}_{n=0}^{\infty}$ are its diagonal terms in the energy basis, and $S$ is the standard (fine-grained) von Neumann entropy. The two entropies coincide if and only if $\rho$ commutes with the Hamiltonian. Obviously, unlike $S$, $S_E$ can increase in a closed system. Decoherence processes eventually  wipe out the off-diagonal components of the density matrix in the energy representation, thus increasing the von Neumann entropy to the level of $S_E$. This makes $S_E$ a convenient tool for the study of non-equilibrium states, especially since unlike the von Neumann entropy, $S_E$ can be conveniently expressed through the Wigner function \cite{Zagoskin2011preprint}, since
\begin{eqnarray}
p_n = 2 (-1)^n \int d\alpha\, d\alpha^* \: e^{-2|\alpha|^2} \: L_n(4|\alpha|^2) \: W(\alpha,\alpha^*) \nonumber \\
= 2 (-1)^n \int\int dx\, dy \: e^{-2(x^2+y^2)}\; L_n(4(x^2+y^2)) \: W(x,y).
\label{eq:pn-general}
\end{eqnarray}
Here $L_n(x)$ is the Laguerre polynomial. We can associate with the additional entropy, $\delta S(s) = S_E(s) - S_E(0)$, of a squeezed state a specific amount of heat,
\begin{equation}
\delta Q(s) = T\:\; \delta S_E(s),
\label{eq:squeezing heat}
\end{equation}
which will be eventually released into the system.

\begin{figure}%
\includegraphics[width=0.8\columnwidth]{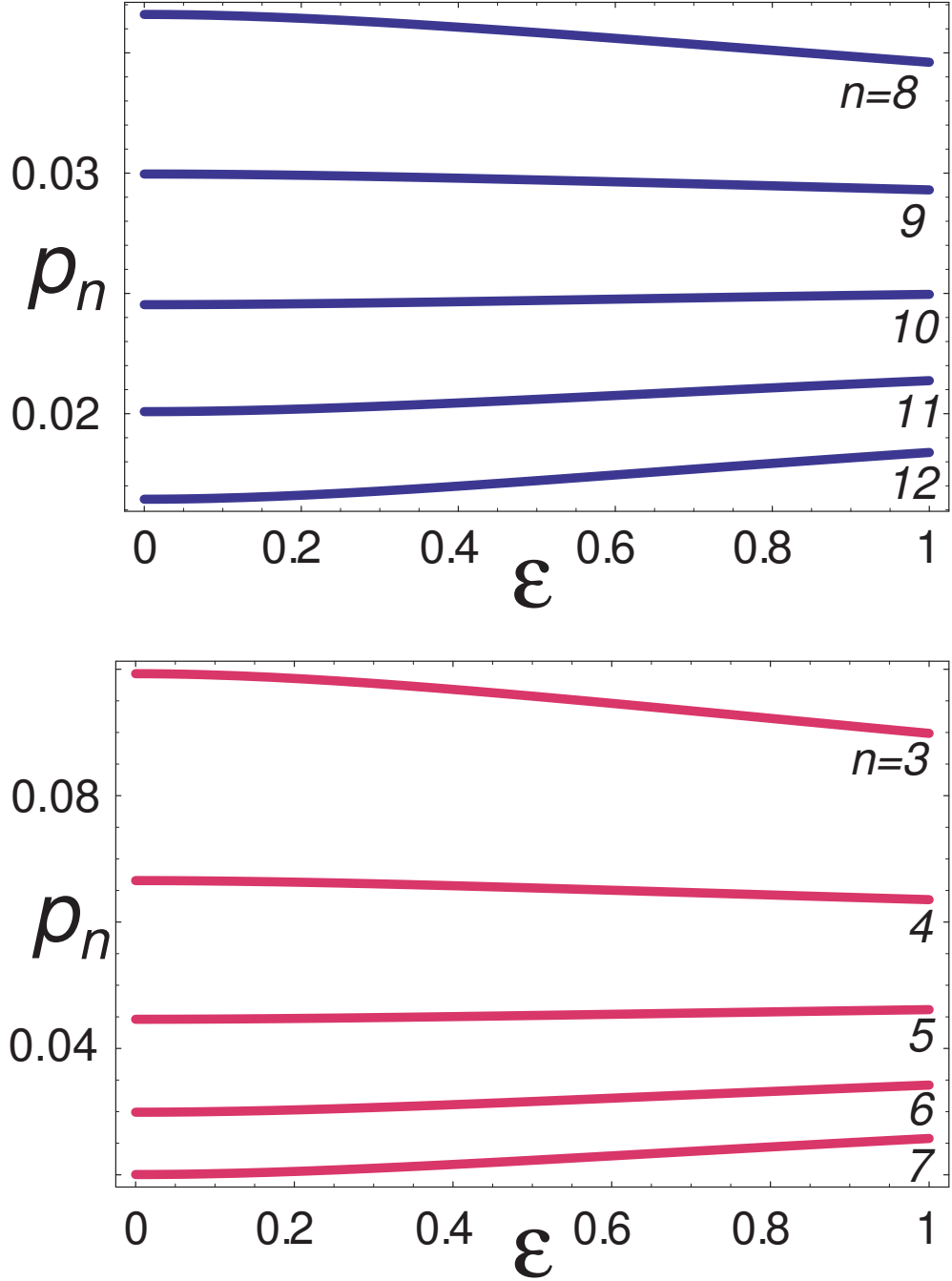}%
\caption{(Color online.) Occupation numbers of the energy levels of a harmonic oscillator in a squeezed thermal state [Eq.~(\ref{eq:pn-final-thermal})] as a function of squeezing $\varepsilon = |s-1|$. (a) $\omega/T = 0.2$,  estimated boundary number $N_0 = 2T/\omega - 1/2 = 9.5$; levels (top to bottom) 8, 9, 10, 11, 12. (b) $\omega/T = 0.4$, $N_0 = 4.5$; levels (top to bottom) 3, 4, 5, 6, 7.}%
\label{fig-1}%
\end{figure}

As an example, consider a squeezed thermal state (\ref{eq:squeezed-Wigner-thermal}), for which an exact solution can be actually found \cite{Zagoskin2011preprint} in terms of the hypergeometric function \cite{Gradstein}:
\begin{equation}
p_n^{\rm th} = \kappa \sum_{q=0}^n C_n^q (-1)^q \left(\frac{2}{1+\frac{\kappa}{s}}\right)^{n+1-q} \left.\right._2{\rm F}_1\left(\frac{1}{2},n+1-q;1;-\frac{\kappa(s-1/s)}{1+\kappa/s}\right),
\label{eq:pn-final-thermal}
\end{equation}
where $\kappa = \tanh(\omega/2T)$. A more convenient approximation, valid for small squeezing, $\kappa(s-1) \ll 1$, is given by
\begin{equation}
p_n \approx \frac{p_n^{\rm eq}}{C_1(\varepsilon)} I_0\left(\kappa\varepsilon\left(\frac{n}{1-\kappa}+\frac{n+1}{1+\kappa}\right)\right) \approx \frac{p_n^{\rm eq}}{C_2(\varepsilon)}  I_0\left(\frac{\omega \varepsilon}{T}(n+1/2)\right).
\label{eq:p-approx}
\end{equation}
Here $p_n^{\rm eq} = (1-\exp[-\omega/T])\exp[-n\omega/T]$ is the equilibrium population of the $n$th energy level of the oscillator, $\varepsilon \equiv |s-1|$ characterizes squeezing, $I_0(z)$ is the modified Bessel function, and the last expression in Eq.~(\ref{eq:p-approx}) is valid  for $\omega/T \ll 1$. The normalization constants $C_{1,2} = 1 + O(\varepsilon^2)$.  The occupation of states with numbers $n < N_0$ will initially decrease, and of those with numbers $n > N_0$ increase with squeezing. Expanding the Bessel function, $I_0(z) = 1 + z^2/4 + \dots$, we see that $d (p_n - p_n^{\rm eq})/d\varepsilon = 0$, if 
\begin{equation}
n = N_0 \equiv \frac{2 T}{\omega} - \frac{1}{2}.
\label{eq:intermediate1}
\end{equation}
This is in  good agreement with the results based on the exact formula in Eq.~(\ref{eq:pn-final-thermal})  (Fig. \ref{fig-1}).

The corresponding change in entropy is given by
\begin{eqnarray}
\delta S_E(\varepsilon) = -\sum_n \delta p_n \ln p_n^{\rm eq} = -\frac{\varepsilon^2}{16} \frac{(\omega/T)^2}{(1-\exp[-\omega/T])^2} \times \nonumber\\
\left\{
\ln(1-e^{-\omega/T}) \left(1 + 6 e^{-\omega/T} + e^{-2\omega/T}\right) - \frac{\omega}{T} \frac{9 e^{-\omega/T} + 14 e^{-2\omega/T} + e^{-3\omega/T}}{1 - e^{-\omega/T}}
\right\}  \nonumber\\
\approx
- \frac{\varepsilon^2}{2} \left\{ 
(1-\frac{\omega}{T}) \ln \frac{\omega}{T} - 3 + 5\frac{\omega}{T} \right\} \:\:\:\:\:\:\:\: \left( \frac{\omega}{T} \ll 1 \right). 
\label{eq:entropy-1}
\end{eqnarray}
With the same accuracy, the equilibrium entropy of an unsqueezed thermal state becomes $$S = -\ln[2\sinh (\omega/2T)] + (\omega/2T) \coth(\omega/2T) \approx - \ln(\omega/T) + 1.$$ Therefore in the leading term in $T/\varepsilon \gg 1$ the squeezing contribution to the entropy is
\begin{equation}
\delta S_E(\varepsilon) \:\: \sim \:\: \frac{\varepsilon^2}{2} S,
\label{eq:entropy-2}
\end{equation}
and the amount of additional heat released into the system due to squeezing will be
\begin{equation}
\delta Q\:\: \sim \:\:T \;\frac{\varepsilon^2}{2} S \: = \: \frac{(s-1)^2}{2} \:T \:S
\label{eq:heat-2}
\end{equation}
per cycle. Detuning between the bus and qubits can be typically 10-30\% \cite{Zagoskin2011}, making the prefactors in Eqs.~(\ref{eq:entropy-2},\ref{eq:heat-2}) as high as 0.25, i.e., the contribution of squeezing to the system entropy becomes comparable to the equilibrium entropy of the device. Of course, this contribution  becomes important only when the system operates near the Landauer erasure limit, with $Q_L = T \ln 2$ \cite{Plenio2001,Maruyama2009}. If a need ever arises to limit its effect, the quadratic dependence of $\delta Q$ on the squeezing parameter indicates that even a slight decrease of the difference between the operating frequencies of the bus will drastically reduce it. One also should avoid a periodic operation of the bus, to exclude the resonant increase of the squeezing rate. 
\\
\\
{\bf CONCLUSION}
\\
\\
We have shown that the operation of a quantum bus will lead to squeezing of its quantum state, and calculated the corresponding additional heat, which will be injected into the system. This contribution is quadratic in the squeezing rate (proportional to the ratio of the working frequencies of the bus) and can become important only in systems operating either near the Landauer erasure limit  or under the conditions when there is a resonant increase of squeezing by repeated bus operation.
\\
\\
{\bf ACKNOWLEDGMENTS}
\\
\\
AZ acknowledges that this publication was made possible through the support of a grant from the John Templeton Foundation; the opinions expressed in this publication are those of the authors and do not necessarily reflect the views of the John Templeton Foundation.  EI  acknowledges partial support from the EU through Solid and IQIT projects. FN is partially supported by the 
LPS, NSA, ARO, NSF grant No. 0726909, JSPS-RFBR contract No. 09-02-92114, Grant-in-Aid for Scientific Research (S), MEXT Kakenhi on Quantum Cybernetics, and the JSPS via its FIRST program.

We thank Prof. J.F. Young for fruitful discussions and Dr. S. Ozdemir for his careful reading of the manuscript and useful suggestions.


\end{document}